\documentclass{article}
\usepackage{spconf,amsmath,graphicx,bbm,array,dcolumn,epsfig,amsmath,amsfonts,yhmath,bm,amssymb,psfrag,color,soul,enumerate,stackengine,cite}
\usepackage[font=footnotesize,skip=7pt]{caption}
\usepackage[font=small,skip=0pt]{subcaption}
\usepackage{tcolorbox}
\usepackage{tikz}

\newcommand*{\myfontb}{\fontfamily{lmr}\selectfont}

\definecolor{lgreen} {RGB}{180,210,100}
\definecolor{ngreen} {RGB}{98,158,31}
\definecolor{dgreen} {RGB}{78,138,21}
\definecolor{MLOWLSgreen} {RGB}{0,140,130}
\definecolor{SDPpurple} {RGB}{191,0,191}
\definecolor{lred}   {RGB}{220,0,0}
\definecolor{nred}   {RGB}{224,0,0}
\definecolor{bred}   {RGB}{200,20,20}
\definecolor{nblue}  {RGB}{28,130,185}
\definecolor{rblue}  {RGB}{28,70,200}
\definecolor{jblue}  {RGB}{20,50,100}
\newcommand*\circled[1]{\tikz[baseline=(char.base)]{
		\node[circle,draw,color=rblue, opacity=0.85,inner sep=0.9pt] (char) {\bf\footnotesize #1};}}

\newcommand {\myvec}[1] {{\mbox{\boldmath $#1$}}}
\newcommand {\mymat}[1]  {{\mbox{\boldmath $#1$}}}

\DeclareMathAlphabet      {\mathbfit}{OML}{cmm}{b}{it}

\AtBeginEnvironment{bmatrix}{\setlength{\arraycolsep}{4pt}}

\newcommand {\A} {\mymat{A}}

\newcommand {\bR} {\mybar{\R}}

\newcommand {\J} {\mymat{J}}

\newcommand {\hC} {\widehat{\C}}

\newcommand {\C} {\mymat{C}}
\newcommand {\D} {\mymat{D}}

\newcommand {\mPhi} {\mymat{\Phi}}
\newcommand {\mPsi} {\mymat{\Psi}}

\newcommand {\uphi} {\myvec{\phi}}
\newcommand {\upsi} {\myvec{\psi}}

\newcommand {\Ep} {\mymat{\mathcal{E}}}

\newcommand {\mSigma} {\mymat{\Sigma}}

\newcommand {\R} {\mymat{R}}
\newcommand {\hR} {\widehat{\R}}
\newcommand {\I} {\mymat{I}}

\newcommand {\ua} {\myvec{a}}

\newcommand {\ubeta} {\myvec{\beta}}

\newcommand {\uc} {\myvec{c}}

\newcommand {\ualpha} {\myvec{\alpha}}
\newcommand {\uv} {\myvec{v}}

\newcommand {\uo} {\myvec{0}}
\newcommand {\us} {\myvec{s}}

\newcommand {\ux} {\myvec{x}}
\newcommand {\ur} {\myvec{r}}
\newcommand {\urho} {\myvec{\rho}}
\newcommand {\uiota} {\myvec{\iota}}
\newcommand {\uy} {\myvec{y}}
\newcommand {\uw} {\myvec{w}}

\newcommand {\utheta} {\myvec{\theta}}
\newcommand {\uhtheta} {\widehat{\myvec{\theta}}}

\newcommand {\ep} {\mathcal{E}}

\newcommand {\Rset} {\mathbb{R}}
\newcommand {\Cset} {\mathbb{C}}

\newcommand {\Eset} {\mathbb{E}}
\newcommand {\Nset} {\mathbb{N}}
\newcommand {\Tr} {\text{\normalfont Tr}}
\newcommand {\Diag} {\text{\normalfont Diag}}
\newcommand {\Ddiag} {\text{\normalfont Ddiag}}

\newcommand {\tps} {\rm{T}}
\newcommand {\her} {\rm{H}}

\newcommand {\calq} {\mathcal{Q}}
\newcommand {\KL} {\mbox{\tiny KL}}
\newcommand {\LS} {\mbox{\tiny LS}}
\newcommand {\sgn} {\mbox{sign}}
\newcommand {\sine} {\mbox{sine}}
\newcommand {\sir} {\mbox{SIR}}
\newcommand {\snr} {\mbox{SNR}}
\newcommand {\kld} {\mbox{\tiny KLD}}
\newcommand {\argmin} {\rm{argmin}}

\makeatletter
\newsavebox\myboxA
\newsavebox\myboxB
\newlength\mylenA

\newcommand*\mybar[2][0.75]{%
	\sbox{\myboxA}{$\m@th#2$}%
	\setbox\myboxB\null
	\ht\myboxB=\ht\myboxA%
	\dp\myboxB=\dp\myboxA%
	\wd\myboxB=#1\wd\myboxA
	\sbox\myboxB{$\m@th\overline{\copy\myboxB}$}
	\setlength\mylenA{\the\wd\myboxA}
	\addtolength\mylenA{-\the\wd\myboxB}%
	\ifdim\wd\myboxB<\wd\myboxA%
	\rlap{\hskip 0.5\mylenA\usebox\myboxB}{\usebox\myboxA}%
	\else
	\hskip -0.5\mylenA\rlap{\usebox\myboxA}{\hskip 0.5\mylenA\usebox\myboxB}%
	\fi}
\makeatother

\title{ENHANCED BLIND CALIBRATION OF UNIFORM LINEAR ARRAYS WITH ONE-BIT QUANTIZATION BY KULLBACK-LEIBLER DIVERGENCE COVARIANCE FITTING}

\name{Amir Weiss$^{\star}$ and Arie Yeredor$^\dagger$}

\address{
\begin{tabular}{cc}
$^{\star}$Dept. of Computer Science and Applied Mathematics & $^\dagger$School of Electrical Engineering\\
Weizmann Institute of Science & Tel-Aviv University\\
amir.weiss@weizmann.ac.il & arie@eng.tau.ac.il
\end{tabular}
}

\begin{document}
\ninept
\maketitle
\setlength{\abovedisplayskip}{5pt}
\setlength{\belowdisplayskip}{5pt}

\begin{abstract}
\small{One-bit quantization has recently become an attractive option for data acquisition in cutting edge applications, due to the increasing demand for low power and higher sampling rates. Subsequently, the rejuvenated one-bit array processing field is now receiving more attention, as ``classical" array processing techniques are adapted / modified accordingly. However, array calibration, often an instrumental preliminary stage in array processing, has so far received little attention in its one-bit form. In this paper, we present a novel solution approach for the {\myfontb\emph{blind}} calibration problem, namely, without using known calibration signals. In order to extract information within the second-order statistics of the quantized measurements, we propose to estimate the unknown sensors' gains and phases offsets according to a Kullback-Leibler Divergence (KLD) covariance fitting criterion. We then provide a quasi-Newton solution algorithm, with a consistent initial estimate, and demonstrate the improved accuracy of our KLD-based estimates in simulations.}
\end{abstract}

\vspace{-0.2cm}
\section{Introduction}\label{sec:intro}
\vspace{-0.2cm}
While high-resolution quantization of analog signals is obviously preferable to low-resolution in terms of accuracy (for signal recovery, feature extraction, etc.), other, practical considerations are often at a higher priority. For example, the power consumption of an Analog-to-Digital Converter (ADC) increases exponentially with the number of bits \cite{walden1999analog}. Other important considerations are the high cost of high-resolution devices \cite{le2005analog}, and their operational sampling rates, which are increasingly becoming insufficient for various applications, such as cognitive
radio and radar \cite{sun2013wideband,stinco2014compressed,smith2016experiments}.

As a remedy, one-bit quantization has been gradually receiving more attention in recent years \cite{choi2016near,gao2017gridless,ren2017one,liu2017one,ameri2019one,huang2019one}. Being essentially the most basic form of quantization, it offers the advantages of low cost and low complexity implementation, which allows for higher sampling rates. These benefits come at the cost of greater loss of information per sample, thus giving rise to the necessity of new, $1$-bit-adapted digital signal processing algorithms.

In the context of passive array processing, one important problem, which nevertheless has so far received little attention in the literature in its $1$-bit form \cite{ramamohan2019blind}, is {\myfontb\emph{calibration}} of the array. This inevitable problem is due to the fact that, in practice, the array sensors generally have different gain and phase responses for various possible reasons (e.g., imperfect manufacturing). Without proper compensation of these relative gain and phase offsets, the accuracy of most (if not all) array-based processing estimation tasks could severely deteriorate. This specific problem is the focus of this work.

While calibration using a known, user-controlled signal is sometimes a viable option, {\myfontb\emph{blind}} calibration is typically more desirable, yet more challenging. In this paper, we consider this challenging problem for Uniform Linear Arrays (ULAs), based on one-bit measurements of narrowband Gaussian signals. We propose a statistically enhanced blind calibration scheme, which exploits ``hidden" Second-Order Statistics (SOS) information via the Kullback-Leibler Divergence (KLD) covariance fitting criterion. The resulting enhanced calibration leads to considerably higher accuracy in subsequent estimation tasks, as demonstrated in a simulation experiment.


\vspace{-0.4cm}
\section{Problem Formulation}\label{sec:problem}
\vspace{-0.2cm}
Consider a ULA of $N$ sensors, each with \textit{unknown} deterministic gain and phase offsets, denoted $\upsi\in\Rset_+^{N \times 1}$ and $\uphi\in[-\pi,\pi)^{N \times 1}$, where $\psi_n$ and $\phi_n$ are the unknown gain and phase offsets of the $n$-th sensor, resp. Further, consider the presence of an unknown number $M$ of unknown, ``far field" narrowband sources, centered around some common carrier frequency with wavelength $\lambda$.

The noisy vector of sampled, baseband-converted signals from all $N$ sensors is given by
\begin{equation}\label{modelequation}
\ur(t)\hspace{-0.03cm}=\hspace{-0.03cm}\mPsi\mPhi\big(\hspace{-0.025cm}\A(\ualpha)\us(t)+\uv(t)\big)+\uw(t)\hspace{-0.03cm}\triangleq\hspace{-0.03cm}\mPsi\mPhi\ux(t)+\uw(t)\hspace{-0.025cm}\in\hspace{-0.025cm}\Cset^{N\times1},
\end{equation}
for all $t\in[T]$, where $[N]\triangleq\{1,\ldots,N\}$ for any $N\in\Nset$, and
\begin{enumerate}[(i)]
	\itemsep0.05em 
	\item $\mPsi\triangleq\Diag(\upsi)\in\Rset_+^{N\times N}$, $\mPhi\triangleq\Diag\left(e^{\jmath\uphi}\right)\in\Cset^{N\times N}$;
	\item $\us(t)\hspace{-0.05cm}\triangleq\hspace{-0.05cm}\left[s_1(t)\,\cdots\,s_M(t)\right]^{\tps}\hspace{-0.05cm}\in\hspace{-0.05cm}\Cset^{M\times1}$ denotes the sources, emitted from unknown azimuth angles $\ualpha\triangleq\left[\alpha_1\,\cdots\,\alpha_M\right]^{\tps}\in\Rset^{M\times1}$;
	\item $\A(\ualpha)\triangleq\left[\ua(\alpha_1)\;\cdots\;\ua(\alpha_M) \right]\in\Cset^{N\times M}$ denotes the nominal array manifold matrix, with the steering vectors $\ua(\alpha_m)\triangleq\left[1\;e^{\jmath \frac{2\pi}{\lambda} d\cos(\alpha_m)}\;\cdots\;e^{\jmath \frac{2\pi}{\lambda}(N-1)d\cos(\alpha_m)}\right]^{\tps}\in\Cset^{N\times1}$ as its columns, where $d\in\Rset_+$ is the inter-element spacing;
	\item $\uv(t)\in\Cset^{N\times1}$ denotes additive, ambient noise or ``interfering" signals, modeled as spatially and temporally independent, identically distributed (i.i.d.) zero-mean circular Complex Normal (CN) \cite{loesch2013cramer} with a covariance matrix $\R_v\triangleq\Eset\left[\uv(t)\uv(t)^{\her}\right]=\sigma_v^2\I_N$, where $\sigma_v^2$ is {\myfontb\emph{unknown}};
	\item $\uw(t)\in\Cset^{N\times1}$ denotes additive internal (e.g., thermal) receiver noise, unaffected by the gain and phase offsets, modeled as spatially and temporally i.i.d.\ zero-mean circular CN with a covariance matrix $\R_w\triangleq\Eset\left[\uw(t)\uw(t)^{\her}\right]=\sigma_w^2\I_N$, where $\sigma_w^2$ is known; and
	\item $\ux(t)$ denotes the signal that would have been received in the absence of gain or phase offsets and internal noise, namely with $\mPsi=\mPhi=\I_N$ and $\sigma_w^2=0$.
\end{enumerate}

We also assume that the sources are mutually uncorrelated. In particular, we assume that $\us(t)$ is a (temporally) i.i.d.\ zero-mean circular CN vector process with an unknown diagonal covariance matrix $\R^s\hspace{-0.075cm}\triangleq\hspace{-0.075cm}\Eset\left[\us(t)\us(t)^{\her}\right]$. In addition, we assume $\us(t), \uv(t)$ and $\uw(t)$ are all mutually statistically independent. As a result we get
\begin{equation}\label{CN_samples}
\ur(t)\sim \mathcal{CN}\left(\uo_N,\R\right), \forall t\in[T],
\end{equation}
where
\begin{align}
\hspace{-0.075cm}\R&\triangleq\Eset\left[\ur(t)\ur(t)^{\her}\right]=\mPsi\mPhi\C\mPhi^*\mPsi+\sigma_w^2\I_N\in\Cset^{N\times N},\label{covariance_of_r}\\
\hspace{-0.075cm}\C\hspace{-0.05cm}&\triangleq\hspace{-0.05cm}\Eset\left[\ux(t)\ux(t)^{\her}\right]\hspace{-0.05cm}=\hspace{-0.05cm}\A(\ualpha)\R^s\A^{\her}(\ualpha)+\sigma_v^2\I_N\in\Cset^{N\times N},\label{covariance_C}
\end{align}
and we have used $\mPsi^{\her}=\mPsi$ and $\mPhi^{\her}=\mPhi^*$. 

In this work, rather than assuming access to the discrete-time signal \eqref{modelequation} measured by an ideal receiver with an $\infty$-bit ADC, we assume access only to a ``coarse" quantized version thereof, obtained by a $1$-bit ADC low-complexity receiver. Specifically, the $1$-bit quantized vector of signals from all $N$ sensors at time $t$ is given by
\begin{equation}\label{onebitmeasuredsignal}
\uy(t)\triangleq\calq\big(\ur(t)\big)\in\left\{e^{\jmath\left(\frac{\pi}{4}+\frac{\pi k}{2}\right)}: 0\leq k\leq3 \right\}^{N\times 1},
\end{equation}
where the complex-valued $1$-bit quantization operator is defined as
\begin{equation*}
\mathcal{Q}(z)\triangleq\frac{1}{\sqrt{2}}\cdot\Big[\sgn\left(\Re\{z\}\right)+\jmath\cdot\sgn\left(\Im\{z\}\right)\Big], \forall z\in\Cset,
\end{equation*}
and, with slight abuse of notations, $\calq(\cdot)$ operates elementwise in \eqref{onebitmeasuredsignal}.

The \textbf{\myfontb \emph{one-bit blind calibration}} problem can now be formulated as follows. {\myfontb \emph{Given the quantized measurements $\left\{\uy(t)\right\}_{t=1}^{T}$ as in \eqref{onebitmeasuredsignal}, estimate the unknown gain and phase offsets $\{\upsi,\uphi\}$.}}

\vspace{-0.4cm}
\section{The Arcsine Law and the Implied SOS-Identifiability}\label{sec:arcsinlaw}
\vspace{-0.3cm}
Before we present our proposed solution approach, it is instructive to study the SOS of the one-bit measurements \eqref{onebitmeasuredsignal}, and the implied identifiability conditions. To this end, following \cite{bar2002doa,liu2017one,ramamohan2019blind}, which all rely on the extension \cite{jacovitti1994estimation} of Van Vleck and Middleton's pioneering work \cite{van1966spectrum}, due to \eqref{CN_samples}, by the {\myfontb\emph{arcsine law}} we have
\begin{equation}\label{defRy}
\R^{y}\triangleq\Eset\left[\uy(t)\uy(t)^{\her}\right]=\frac{2}{\pi}\sine^{-1}\left(\bR\right)\in\Cset^{N\times N},
\end{equation}
where $\bR$ is the normalized covariance matrix of $\ur(t)$, defined as
\begin{equation}\label{defRbar}
\bR\triangleq\Ddiag(\R)^{-\frac{1}{2}}\R\hspace{0.05cm}\Ddiag(\R)^{-\frac{1}{2}}\triangleq\D^{-\frac{1}{2}}\R\D^{-\frac{1}{2}},
\end{equation}
where $\Ddiag(\R)$ denotes a diagonal matrix with the same diagonal elements as $\R$, such that $\left|\mybar{R}_{ij}\right|\leq1$ for all $i,j\in[N]$, $\mybar{R}_{ii}=1$ for all $i\in[N]$, and $\sine^{-1}(z)\triangleq\sin^{-1}(\Re\{z\})+\jmath\cdot\sin^{-1}(\Im\{z\})$ operates elementwise.

Since the array manifold matrix of a ULA is a Vandermonde matrix (e.g., \cite{petersen2008matrix}) and all the signals involved are uncorrelated, the covariance matrix $\C$ \eqref{covariance_C} is a Toeplitz matrix \cite{gray2006toeplitz}, thus we denote $C_{ij}\triangleq c_{|i-j|+1}$, so the $(i,j)$-th element of $\bR$ can be expressed as
\begin{equation*}\label{elementofRy}
\mybar{R}_{ij}=\frac{R_{ij}}{\sqrt{D_{ii}D_{jj}}}=\frac{\psi_i\psi_jc_{|i-j|+1}e^{\jmath(\phi_i-\phi_j)}+\delta_{ij}\sigma_w^2}{\sqrt{(\psi^2_ic_1+\sigma_w^2)(\psi^2_jc_1+\sigma_w^2)}},
\end{equation*}
where $\delta_{ij}$ denotes the Kronecker delta of $i,j\in\Nset$. Indeed, it is readily verified that $\mybar{R}_{ii}=1$ for all $i\in[N]$. In addition, since both $\R^s$ and $\sigma_v^2$ are unknown, and since all $\{\mybar{R}_{ij}\}$ are related to $\{\psi_i\}$ and $\{c_{|i-j|+1}\}$ by the products $\{\psi_i\psi_jc_{|i-j|+1}\}$ {\myfontb\emph{only}}, we can set w.l.o.g.\ $c_{1}=1$. We will shortly see that the (unknown) Signal-to-Interference Ratios $\sir_m\triangleq R^s_{mm}/\sigma_v^2$ are immaterial to our blind calibration problem. This is in stark contrast to the Signal-to-Noise Ratios, $\snr_n\triangleq\psi_n^2/\sigma_w^2$, which, after setting $c_1=1$, are now expressed via the sensors' gains $\{\psi_n\}$, and are also unknown.

To see this, observe that in the limit of high SNR we have
\begin{equation*}\label{limithighsnr}
\lim_{\sigma_w^2\to0}\mybar{R}_{ij}=\frac{\psi_i\psi_jc_{|i-j|+1}e^{\jmath(\phi_i-\phi_j)}}{\sqrt{\psi^2_i\psi^2_j}}=c_{|i-j|+1}e^{\jmath(\phi_i-\phi_j)},
\end{equation*}
namely a complete loss of information regarding the sensors' gains. Note that \cite{ramamohan2019blind} addresses (only) this particular case in the context of blind calibration for sparse arrays, consequently only calibrating the sensors' phases. Our signal model \eqref{modelequation} therefore offers an extended framework (\textit{cf}.\ \cite{ramamohan2019blind}, Eq.\ (1)), which allows blind calibration of the sensors' gains as well for any finite (even if large) SNR, provided that the sample size is sufficiently large. Be that as it may, from a practical point of view, at high SNR the noiseless model still leads to ``good" solutions for phase calibration (only).

Lastly, recall that even with an ideal $\infty$-bit ADC, and therefore in our case as well, at least two phases and one gain references are required for calibration of the array in this blind scenario (e.g., \cite{astely1999spatial}).

\vspace{-0.6cm}
\section{One-Bit Blind Calibration}\label{sec:blindcalibrationOWLS}
\vspace{-0.3cm}
In light of the discussion above, w.l.o.g., we assume from now on
\begin{equation}\label{refequations}
\psi_1=1, \; \phi_1=\phi_2=0, \; c_1=1.
\end{equation}
With these assumptions, we define the underlying vector of unknown parameters, which fully determines $\R^y$ with $\sigma_w^2$ and \eqref{refequations}, as
\begin{equation*}\label{thetadef}
\utheta\triangleq\left[\upsi_{[2]}^{\tps}\;\,\uphi_{[3]}^{\tps}\;\,\urho_{[2]}^{\tps}\;\,\uiota_{[2]}^{\tps} \right]^{\tps}\in\Rset^{K_{\theta}\times1},
\end{equation*}
where we use the notation $\ua_{[k]}\hspace{-0.05cm}\triangleq\hspace{-0.05cm}[a_k\;a_{k+1}\,\cdots\,a_{N}]^{\tps}\hspace{-0.08cm}\in\Rset^{(N-k+1)\times1}$ for any $\ua\in\Rset^{N\times 1}$ and $k\leq N$, and where $\uc\triangleq\urho+\jmath\cdot\uiota\in\Cset^{N\times 1}$, such that $\rho_1=1$ and $\iota_1=0$ from \eqref{refequations}, and $K_{\theta}\triangleq4N-5$.

Many of the previously proposed methods for blind calibration are based on the fact that, theoretically, $\hR^y\triangleq\frac{1}{T}\sum_{t=1}^{T}{\uy(t)\uy(t)^{\her}}\in\Cset^{N\times N}$ can be made arbitrarily close to $\R^y$ by increasing (appropriately) the sample size $T$. However, in practice, the available sample size is always limited and is oftentimes fixed. Hence, rather than relying on the coarse approximation $\hR^y\approx\R^y$, which typically leads to coarse, sub-optimal estimates of $\{\upsi,\uphi\}$, we propose a refined approach, which obtains more accurate estimates of $\upsi$ and $\uphi$ by accounting for information regarding the estimation errors in $\hR^y$, which are generally correlated \cite{weiss2020asymptotically}.

Specifically, for any finite sample size $T$, we have
\begin{equation*}\label{Restimate}
\hR^y\triangleq\R^y+\Ep\;\Longrightarrow\;\widehat{R}^y_{ij}=R^y_{ij}+\ep_{ij}, \; \forall i,j\in[N],
\end{equation*}
where $\{\ep_{ij}\}$ denote the estimation errors in the estimation of $\{R^y_{ij}\}$. Using \eqref{defRy}, it is readily seen that while $\sine\left(\frac{\pi}{2}R^y_{ij}\right)=\mybar{R}_{ij}$,
\begin{equation*}\label{arcsinRestimate}
\sine\left(\frac{\pi}{2}\widehat{R}^y_{ij}\right)=\sine\left(\sine^{-1}(\mybar{R}_{ij})+\frac{\pi}{2}\ep_{ij}\right)\triangleq \mybar{R}_{ij}+\widetilde{\ep}_{ij},
\end{equation*}
where $\sine(z)\triangleq\sin(\Re\{z\})+\jmath\cdot\sin(\Im\{z\})$, and $\{\widetilde{\ep}_{ij}\}$ are the implied ``transformed" estimation errors. A direct, explicit SOS-based analytic characterization of $\widetilde{\ep}_{ij}$ is rather involved, so we seek an estimate which would {\myfontb\emph{implicitly}} account for correlations among $\{\ep_{ij}\}$. This leads us to the following solution approach.

\vspace{-0.5cm}
\subsection{Estimation by KLD Covariance Fitting}\label{subsec:KLDcovfit}
\vspace{-0.2cm}
We propose to use the KLD of two zero-mean circular CN distributions as a covariance fitting criterion, in order to implicitly extract more information contained is the SOS. Since the zero-mean CN distribution is fully characterized by SOS, note that the KLD of two zero-mean multivariate CN distributions serves as a plausible criterion for {\myfontb\emph{consistent}} covariance matrix estimation. Indeed, from Gibbs' inequality, the KLD is always non-negative (e.g., \cite{mackay2003information}), hence
\begin{gather}
D^{\mathcal{CN}}_{\KL}\left(\widehat{\mSigma},\mSigma\right)\triangleq\log\left(\frac{\det\mSigma}{\det\widehat{\mSigma}}\right) + \Tr\left(\widehat{\mSigma}\mSigma^{-1}\right) - N\ge 0,\label{KLDifffcond}\\
D^{\mathcal{CN}}_{\KL}\left(\widehat{\mSigma},\mSigma\right)=0 \iff \widehat{\mSigma}=\mSigma\in\Cset^{N\times N}.
\end{gather}
Moreover, it can be easily shown that minimization of \eqref{KLDifffcond} is equivalent to maximization of the likelihood function for CN-distributed measurements (e.g., \cite{weiss2020blind}, Eq. (60)), which, in turn, is asymptotically equivalent to minimization of the optimally weighted nonlinear Least Squares (LS) objective function (see \cite{weiss2020blind}, Eq.\ (62), and Appendix E therein for the full details). Specifically, we propose
\begin{equation}\label{KLDestimate}
\uhtheta_{\kld}\triangleq \underset{{\text{\boldmath $\theta$}}\in\Rset^{K_{\theta}\times 1}}{\argmin} D^{\mathcal{CN}}_{\KL}\left(\hR^y,\R^y(\utheta)\right).
\end{equation}
Thus, due to all of the above, the proposed estimate \eqref{KLDestimate}, from which the KLD-based estimates of $\upsi_{[2]}$ and $\uphi_{[3]}$ are readily extracted, implicitly takes into account cross-correlations between all the estimation errors $\{\ep_{ij}\}$ and utilizes ``hidden" SOS information, as desired.

\vspace{-0.4cm}
\subsection{Computation of the Proposed KLD-based Estimate}\label{subsec:computationofest}
\vspace{-0.15cm}
As mentioned above, since $\uhtheta_{\kld}$ is equivalent to the maximum likelihood estimate for CN-distributed measurements, we propose to use the respective {\myfontb\emph{Fisher's Scoring Algorithm}} (FSA, e.g., \cite{jennrich1976newton}), pretending that $\{\uy(t)\}_{t=1}^T$ are CN-distributed. Of course, in our case the FSA would merely serve as a quasi-Newton algorithm (e.g., \cite{bonnans2006numerical}), since $\{\uy(t)\}_{t=1}^{T}$ are obviously far from being CN-distributed. 

The update equation of the FSA for the $k$-th iteration is given by
\begin{equation}\label{FSAupdatequation}
\uhtheta^{(k)} = \uhtheta^{(k-1)} + \J^{-1}\left(\uhtheta^{(k-1)}\right)\left.\nabla_{\bm{\theta}}\mathcal{L}\right|_{\scriptsize{\utheta}=\uhtheta^{(k-1)}},
\end{equation}
where $\uhtheta^{(k)}$ denotes the estimate of $\utheta$ in the $k$-th iteration,
\begin{equation}\label{loglikelihood}
\mathcal{L}(\utheta)\triangleq -T\cdot D^{\mathcal{CN}}_{\KL}\left(\hR^y,\R^y(\utheta)\right) + b,
\end{equation}
is the associated log-likelihood function (where $b\in\Rset$ is constant w.r.t.\ $\utheta$), and $\J(\utheta)\in\Rset^{K_{\theta}\times K_{\theta}}$ is the associated Fisher Information Matrix (FIM) under the (false) assumption that $\{\uy(t)\}_{t=1}^T$ are CN, given by \cite{collier2005fisher} (for all $i,j\in[K_{\theta}]$)
\begin{equation}\label{FIM_CN_element}
J_{ij}(\utheta)=T\cdot{\Tr\left(\left(\R^y\right)^{-1}\left(\tfrac{\partial}{\partial \theta_i}\R^y\right)\left(\R^y\right)^{-1}\left(\tfrac{\partial}{\partial \theta_j}\R^y\right)\right)}.
\end{equation}
Hence, in order to carry out the iterations \eqref{FSAupdatequation} such that they will successfully converge to $\uhtheta_{\kld}$, three ingredients are required: a sufficiently ``good" initial solution $\uhtheta^{(0)}$, and closed-form expressions of the FIM and the score function, i.e., $\J(\utheta)$ and $\nabla_{\bm{\theta}}\mathcal{L}\in\Rset^{K_{\theta}\times 1}$, resp.

Starting with the elements of $\J(\utheta)$, and denoting $\mybar{R}_{mn}\triangleq\mybar{R}^{\Re}_{mn}+\jmath\cdot\mybar{R}^{\Im}_{mn}$ for brevity, by the chain rule, we have for every $\theta_i$
\begin{gather}\label{derivativeofRy}
\hspace{-0.15cm}\frac{\partial R^y_{mn}}{\partial\theta_i}\hspace{-0.025cm}=\hspace{-0.025cm}\frac{2/\pi}{\sqrt{1-\left(\mybar{R}^{\Re}_{mn}\right)^2}}\frac{\partial\mybar{R}^{\Re}_{mn}}{\partial\theta_i}\hspace{-0.05cm}+\hspace{-0.05cm}\frac{\jmath\cdot2/\pi}{\sqrt{1-\left(\mybar{R}^{\Im}_{mn}\right)^2}}\frac{\partial\mybar{R}^{\Im}_{mn}}{\partial\theta_i},
\end{gather}
where we have used $\frac{\partial}{\partial x}\sin^{-1}(x)=\left(1-x^2\right)^{-\frac{1}{2}}$. Given \eqref{derivativeofRy}, all $\{\frac{\partial}{\partial\theta_i}\R^y\}$ can now be easily computed using straightforward derivatives, omitted due to lack of space. As a representative example,
\begin{align*}
&\frac{\partial\mybar{R}^{\Re}_{mn}}{\partial\phi_k}\hspace{-0.05cm}=\hspace{-0.05cm}\left(\rho_{|m-n|+1}\sin(\phi_m\hspace{-0.05cm}-\hspace{-0.05cm}\phi_n)+\iota_{|m-n|+1}\cos(\phi_m\hspace{-0.05cm}-\hspace{-0.05cm}\phi_n)\right)\cdot\\
&\frac{\psi_m\psi_n}{\sqrt{D_{mm}D_{nn}}}(\delta_{km}-\delta_{kn}), \; \forall m\neq n\in[N], \forall (k-2)\in[N-2].
\end{align*}
With $\{\frac{\partial}{\partial\theta_i}\R^y\}_{i=1}^{K_{\theta}}$, we readily obtain closed-form expressions for all of \eqref{FIM_CN_element}. As for $\nabla_{\bm{\theta}}\mathcal{L}$, using the chain rule once again, we have
\begin{equation*}\label{score_wrt_theta}
\frac{\partial\mathcal{L}(\utheta)}{\partial\theta_i}=\sum_{m,n=1}^{N}{\frac{\partial\mathcal{L}\left(\utheta\right)}{\partial R^y_{mn}}\frac{\partial R^y_{mn}}{\partial \theta_i}}=\Tr\left(\big(\nabla_{\bm{R}^y}\mathcal{L}\big)\left(\tfrac{\partial}{\partial\theta_i}{\R^y}\right)^{\tps}\right),
\end{equation*}
for all $i \in[K_{\theta}]$, where
\begin{equation}\label{score_wrt_covmatR}
\nabla_{\bm{R}^y}\mathcal{L}=-T\cdot\left[\left(\R^y\right)^{-1}\left(\I_{N}-\hR^y\left(\R^y\right)^{-1}\right)\right]^{\tps}\in\Cset^{N\times N}.
\end{equation}
Thus, with \eqref{score_wrt_covmatR} and the already obtained $\{\frac{\partial}{\partial\theta_i}\R^y\}_{i=1}^{K_{\theta}}$, we now have a closed-form expression for the score function $\nabla_{\bm{\theta}}\mathcal{L}$ as well.

The third ingredient---a ``good" initial solution to the iterative procedure \eqref{FSAupdatequation}---is instrumental from a practical point of view. Indeed, generally, the FSA is likely to converge to \eqref{KLDestimate} when $\uhtheta^{(0)}$  is ``close" enough to $\widehat{\utheta}_{\kld}$, which is the global maximizer of \eqref{loglikelihood}.

For this, we propose to use Paulraj and Kailath's (P-K's) classical LS approach \cite{paulraj1985direction}. Despite ignoring the estimation errors $\{\ep_{ij}\}$, P-K's approach yields simple closed-form expressions for consistent estimates of $\{\upsi_{[2]},\uphi_{[3]}\}$, based on the Toeplitz structure of $\C$, which can still be exploited under one-bit quantization. Indeed, $\widehat{\uphi}_{[3],\LS}$ can be obtained just as described in \cite{paulraj1985direction}, using $\widehat{\bR}\triangleq\sine\left(\frac{\pi}{2}\hR^y\right)$. Further, $\widehat{\upsi}_{[2],\LS}$ can be obtained as well with the following, relatively simple modifications. Since $\C$ is Toeplitz, due to \eqref{covariance_of_r} and \eqref{defRbar}, we have (\textit{cf}. \cite{paulraj1985direction}, Eq. (11))
\begin{equation*}\label{reallogofRbar}
\mu_{ijk\ell}\triangleq\log\left(\frac{|\mybar{R}_{ij}|}{|\mybar{R}_{k\ell}|}\right)=\beta_i-\beta_j-\beta_k+\beta_{\ell},
\end{equation*}
for any four indices satisfying $i-j=k-\ell\neq0$, where
\begin{equation}\label{defbeta}
\beta_n\triangleq\log\left(\frac{\psi_n}{\sqrt{\psi_n^2+\sigma_w^2}}\right) \iff \psi_n=\sqrt{\frac{e^{2\beta_n}\sigma_w^2}{1-e^{2\beta_n}}}, \, \forall n\in[N].
\end{equation}
Thus, recalling that $\sigma_w^2$ and $\psi_1$ are known, which means that $\beta_1$ is known, the LS estimate $\widehat{\ubeta}_{[2],\LS}$ based on $\widehat{\bR}$ is obtained via \cite{paulraj1985direction}, Eq. (13) (excluding equations associated with the diagonal elements). Given $\widehat{\ubeta}_{[2],\LS}$, the estimates $\widehat{\upsi}_{[2],\LS}$ are readily computed from \eqref{defbeta}.

\begin{figure}[b]
	\centering
	\begin{subfigure}[b]{0.2\textwidth}
		\includegraphics[width=\textwidth]{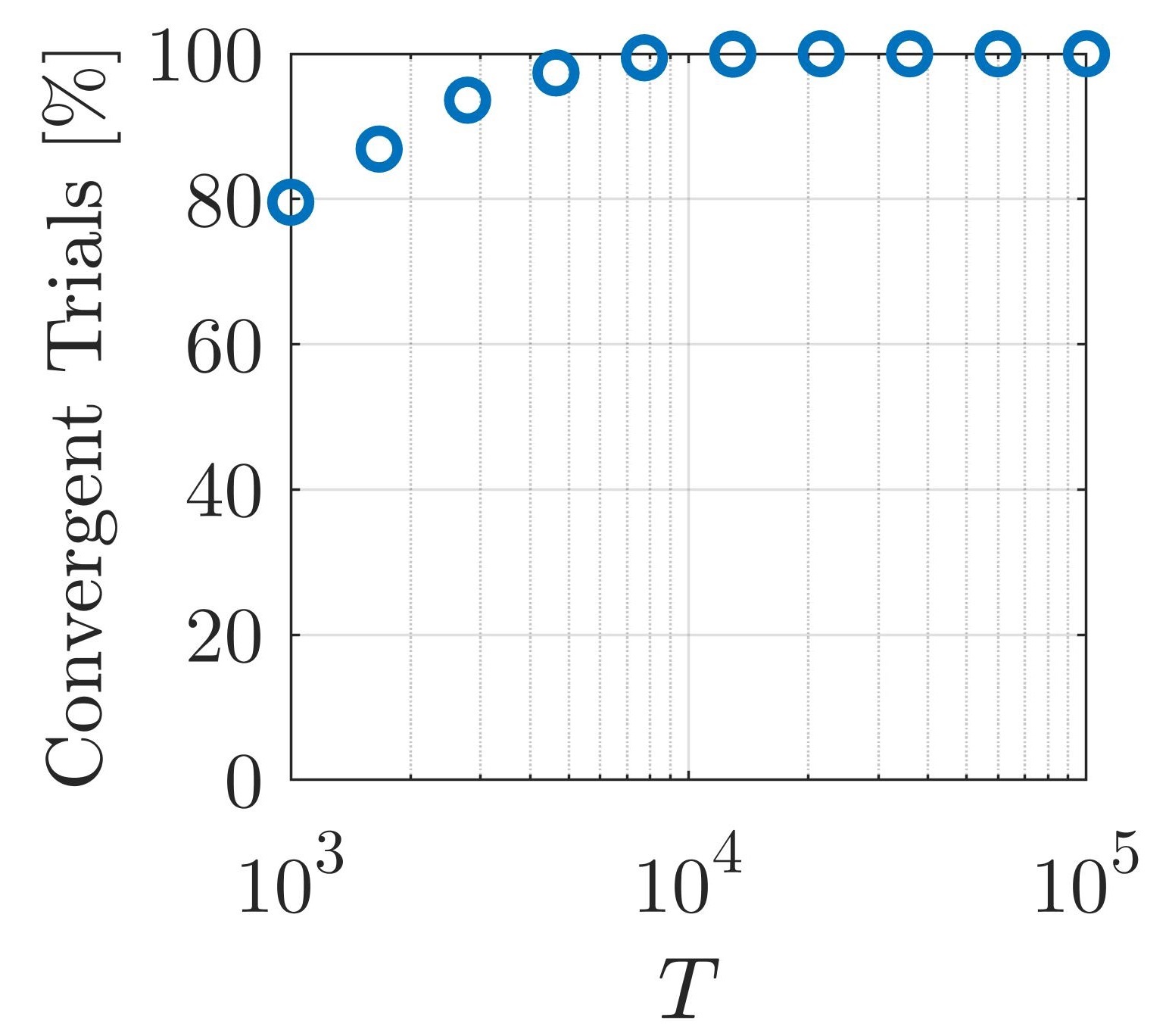}
		\caption{}
		\label{fig:convergence_plot_a}
	\end{subfigure}%
	~~~~
	\begin{subfigure}[b]{0.2\textwidth}
		\includegraphics[width=\textwidth]{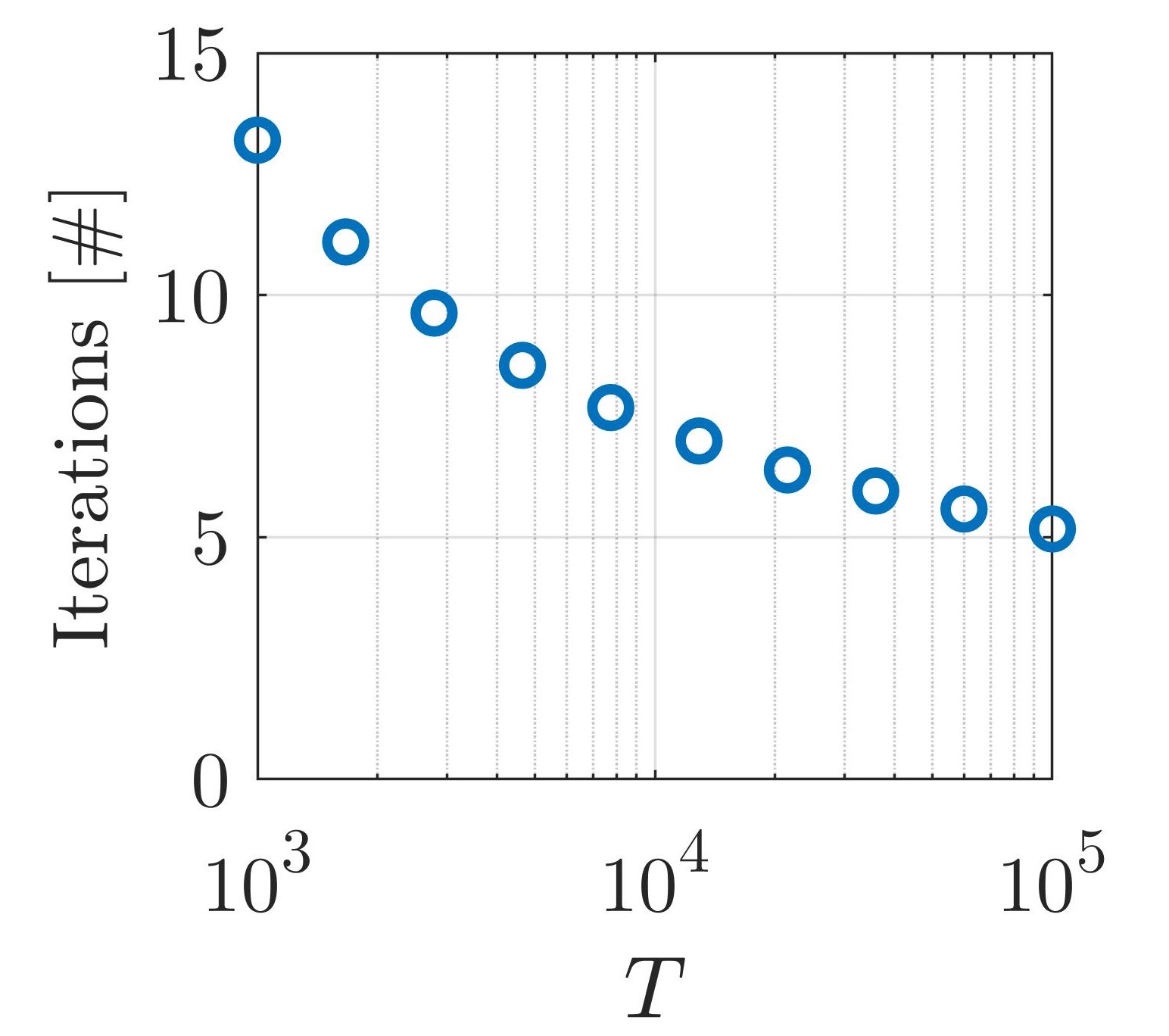}
		\caption{}
		\label{fig:convergence_plot_b}
	\end{subfigure}
	\vspace{-0.1cm}
	\caption{(a) Percentage of convergent FSA \eqref{FSAupdatequation} trials; (b) Upon convergence, average number of iterations to convergence, both vs.\ $T$, for $\uhtheta^{(0)}=\uhtheta_{\LS}$.} 
	\label{fig:convergence_plot}
	\vspace{-0.35cm}
\end{figure}
\begin{figure*}[t]
	\centering
	\includegraphics[width=.95\textwidth]{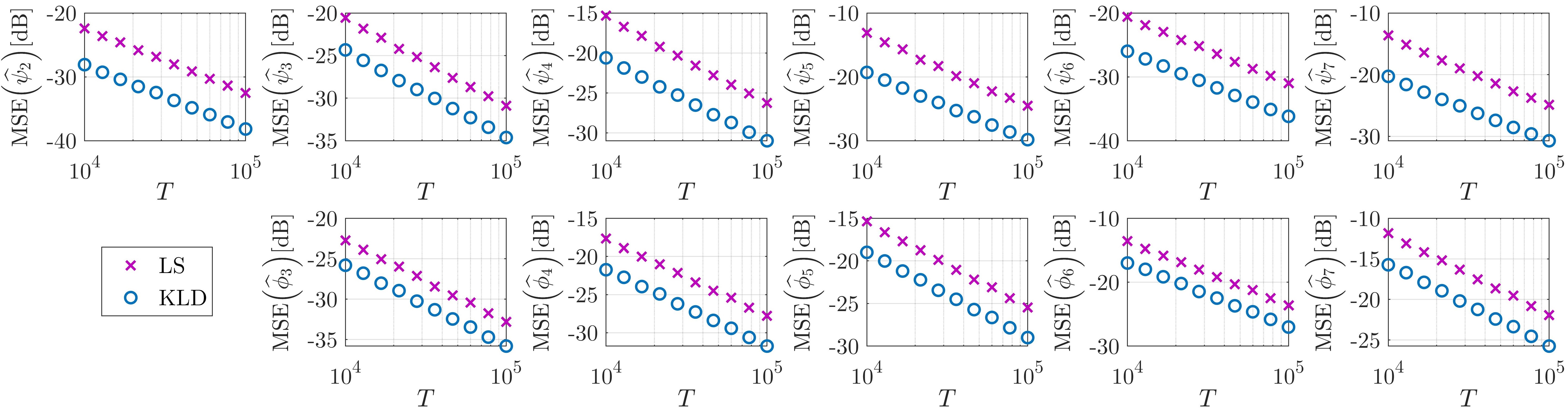}
	\caption{MSEs of the gains and phases offsets estimates vs.\ $T$. Though both the LS- and KLD-based estimates exhibit a consistency trend, the KLD are superior.}
	\label{fig:MSE_vs_T}
	\vspace{-0.4cm}
\end{figure*}
\begin{figure}[t]
	\centering
	\includegraphics[width=0.375\textwidth]{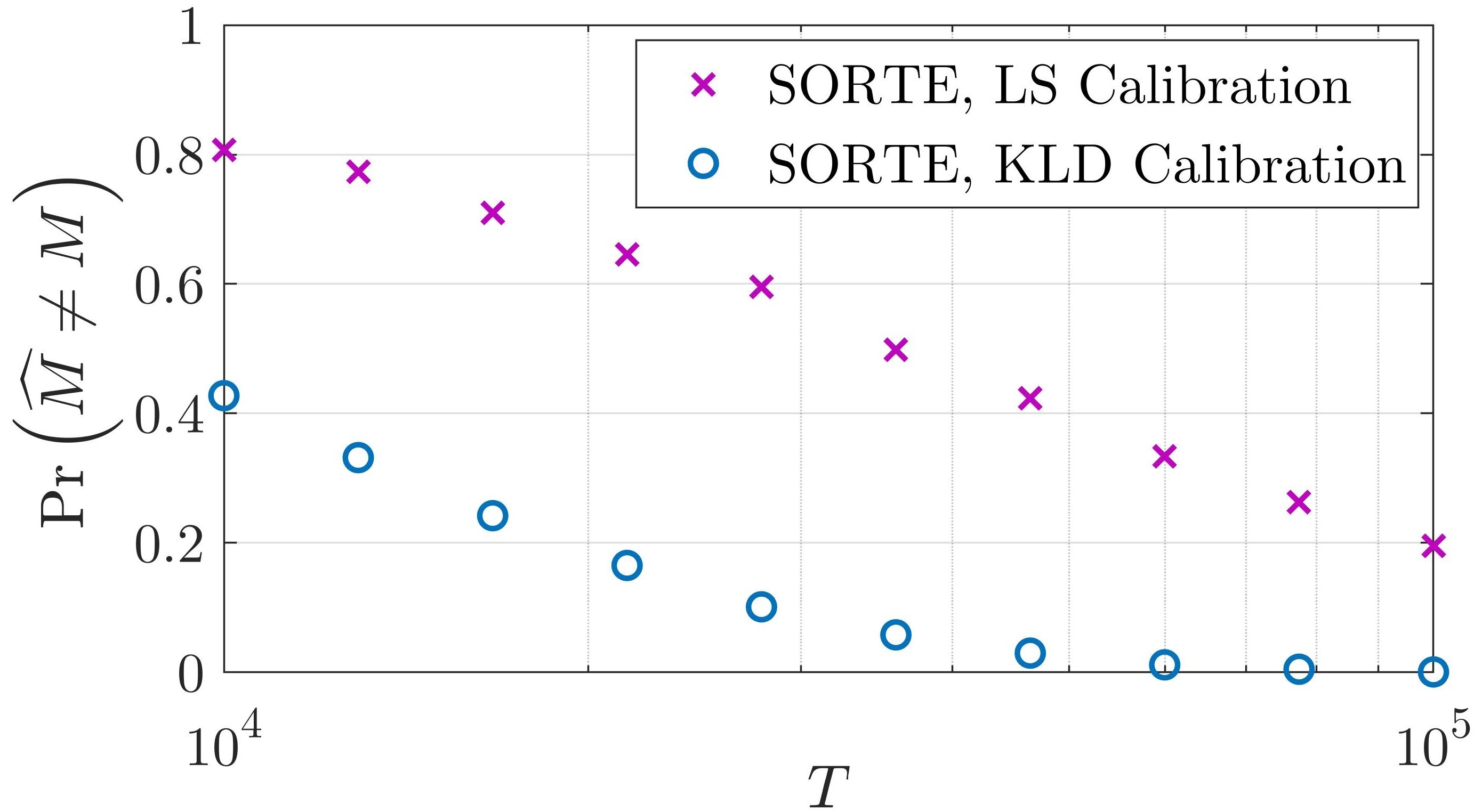}
	\caption{Empirical probability of error in detecting the number of sources vs.\ $T$. Clearly, the KLD-based enhanced calibration significantly improves the accuracy in this post-calibration covariance-based estimation task.}
	\label{fig:error_probability_SORTE}
	\vspace{-0.5cm}
\end{figure}

Once $\widehat{\upsi}_{[2],\LS}$ and $\widehat{\uphi}_{[3],\LS}$ are obtained, we define the diagonal matrix $\widehat{\D}_{\LS}$ with diagonal entires $\widehat{D}_{nn,\LS}\triangleq\widehat{\psi}_{n,\LS}^2+\sigma_w^2$, with which we further define $\widehat{\R}_{\LS}\triangleq\widehat{\D}_{\LS}^{\frac{1}{2}}\hspace{0.02cm}\widehat{\bR}\hspace{0.02cm}\widehat{\D}_{\LS}^{\frac{1}{2}}$ based on \eqref{defRbar}, and (based on \eqref{covariance_of_r})
\begin{gather*}\label{LSestofC}
\hC_{\LS}\triangleq\widehat{\mPhi}_{\LS}^{*}\widehat{\mPsi}_{\LS}^{-1}\left(\widehat{\R}_{\LS}-\sigma_w^2\I_N\right)\widehat{\mPsi}_{\LS}^{-1}\widehat{\mPhi}_{\LS},\\
\widehat{c}_{|i-j|+1,\LS}\triangleq \widehat{C}_{ij,\LS} \; \Longrightarrow \; \widehat{\uc}_{\LS}\triangleq\widehat{\urho}_{\LS}+\jmath\cdot\widehat{\uiota}_{\LS},
\end{gather*}
where $\widehat{\mPsi}_{\LS}\triangleq\Diag\left(\widehat{\upsi}_{\LS}\right), \widehat{\mPhi}_{\LS}\triangleq\Diag\left(\widehat{\uphi}_{\LS}\right), \widehat{\rho}_{1,\LS}=1$ and $\widehat{\iota}_{1,\LS}=0$. Collecting all the respective LS-based estimates into a $K_{\theta}$-dimensional vector, we now have
\begin{equation}\label{thetaLSest}
\widehat{\utheta}_{\LS}\triangleq\left[\widehat{\upsi}_{[2],\LS}^{\tps}\;\widehat{\uphi}_{[3],\LS}^{\tps}\;\widehat{\urho}_{[2],\LS}^{\tps}\;\widehat{\uiota}_{[2],\LS}^{\tps} \right]^{\tps}\triangleq\uhtheta^{(0)}\in\Rset^{K_{\theta}\times1},
\end{equation}
which is a consistent estimate of $\utheta$ (from the consistency of $\hR^y$), thus serving as a ``good" initial solution, presumably ``close" to $\uhtheta_{\kld}$.

The complete proposed blind calibration scheme is as follows:
\tcbset{colframe=gray!95!blue,size=small,width=0.49\textwidth,arc=2.1mm,outer arc=1mm}
\begin{tcolorbox}[upperbox=visible,colback=white]
	\textbf{\underline{KLD-based One-Bit Blind Calibration}:} Given $\{\uy(t)\}_{t=1}^T$,
	\begin{enumerate}
	\itemsep0.025em 
	\item[\circled{1}] Compute $\hR^y=\frac{1}{T}\sum_{t=1}^{T}\uy(t)\uy(t)^{\her}$;
	\item[\circled{2}] Compute and construct $\uhtheta_{\LS}$ as in \eqref{thetaLSest};
	\item[\circled{3}] Set $\uhtheta^{(0)}=\uhtheta_{\LS}$, and iterate \eqref{FSAupdatequation} until convergence (e.g., until $\left\|\uhtheta^{(k)}-\uhtheta^{(k-1)}\right\|_2<\epsilon$, for some ``small" $\epsilon\in\Rset_+$);
	\item[\circled{4}] Calibrate the array using $\widehat{\upsi}_{\kld}\triangleq\left[1 \;\, \widehat{\upsi}^{\tps}_{[2],\kld}\right]^{\tps}$ and $\widehat{\uphi}_{\kld}\triangleq\left[0 \;\, 0 \;\, \widehat{\uphi}^{\tps}_{[3],\kld}\right]^{\tps}$.
	\end{enumerate}
\end{tcolorbox}\vspace{-0.1cm}

We note that even if access to the pre-quantized signals is restricted, so that $\widehat{\upsi}_{\kld}$ and $\widehat{\uphi}_{\kld}$ cannot be compensated for {\myfontb\emph{prior}} to the one-bit quantization, our method nevertheless provides $\hC_{\kld}$---the KLD-based estimate of $\C$---which is readily constructed from $\widehat{\urho}_{\kld}\triangleq[1 \;\, \widehat{\urho}_{[2],\kld}^{\tps}]^{\tps}$ and $\widehat{\uiota}_{\kld}\triangleq[0 \;\, \widehat{\uiota}_{[2],\kld}]^{\tps}$, and can be used for subsequent post-calibration covariance-based estimation tasks. $\hC_{\kld}$ essentially serves as the outcome of implicit KLD-based blind calibration, applied directly to the one-bit quantized measurements.

\vspace{-0.3cm}
\section{Simulation Results}\label{sec:results}
\vspace{-0.2cm}

We consider model \eqref{modelequation} with one-bit measurements \eqref{onebitmeasuredsignal}, a half wavelength inter-element spacing $d=\lambda/2$ ULA consisting of $N=7$ elements, and $M=4$ equal power, zero-mean sources arriving from angles $\alpha=[45^\circ\;52^\circ\;9^\circ\;78^\circ]^{\tps}$. The sensors' gains and phases were set to $\upsi=[1\;0.7\;0.9\;1.1\;1.2\;0.8\;1.3]^{\tps}$ and $\uphi=[0^\circ\,0^\circ\,5^\circ\,11^\circ\,-\hspace{-0.05cm}8^\circ\,4^\circ\,10^\circ]^{\tps}$, resp., where w.l.o.g.\ we assume $\psi_1, \phi_1$ and $\phi_2$ are known references. The sources' (equal) power was set such that $\{\sir_m=10\}_{m=1}^4$ and $c_1=1$, and $\sigma_w^2=1$. All empirical results were obtained by averaging $10^4$ independent trials.

Figs.\ \ref{fig:convergence_plot_a} and \ref{fig:convergence_plot_b} present the empirical percentage of convergent trials and number of iterations to convergence, resp., of FSA vs. $T$, when initialized by $\uhtheta_{\LS}$. Here, we say that the iterations \eqref{FSAupdatequation} converge when the $\ell^2$-norm of the step size falls below $\epsilon=10^{-7}$ before reaching the maximum number of allowed iteration $K_{\text{max}}=100$. It is clearly demonstrated that when the sample size is sufficiently large, the iterations converge with high probability, which is due to the consistency of $\uhtheta_{\LS}$. We also emphasize that although the nominal values of $T$ considered here may seem typically large, the respective number of {\myfontb\emph{bits per unkonwn}} is both realistic and practical owing to the high operational rates of low-complexity $1$-bit ADCs.

Next, we turn to evaluate the MSEs of the proposed $\widehat{\upsi}_{\kld}$ and $\widehat{\uphi}_{\kld}$, and their more simple LS-based alternatives $\widehat{\upsi}_{\LS}$ and $\widehat{\uphi}_{\LS}$, presented in Fig.\ \ref{fig:MSE_vs_T} vs.\ $T$. For a pure (and fair) comparison of the MSEs (regardless of convergence issues), the average MSEs of the KLD-based estimates are based only on convergent FSA trials. As seen, the proposed estimates exhibit enhanced performance, reducing the MSEs by up to $\sim\hspace{-0.05cm}6$[dB]. For practical considerations, whenever FSA does not converge, one can resort to calibration by $\widehat{\upsi}_{\LS}$ and $\widehat{\uphi}_{\LS}$.

Lastly, we demonstrate the implications of our enhanced solution by a subsequent post-calibration task---blind determination of the number of sources (e.g., \cite{weiss2019blindnumberofsources}). To this end, we use $\hC_{\kld}$ and $\hC_{\LS}$ as the post-calibration estimated covariance matrix of the received signal $\ux(t)$, where the goal now is to detect the number of sources $M$, recalling $\sigma_v^2$ is unknown. We use the second order statistic of the eigenvalues (SORTE) algorithm \cite{he2010detecting}, which requires only the eigenvalues of the sample covariance matrix (of a calibrated array). Fig.\ \ref{fig:error_probability_SORTE}, presenting the empirical probability of error in detecting $M$ vs.\ $T$, well demonstrates the considerable performance improvement, a direct implication of the proposed KLD-based enhanced calibration.
\vspace{-0.7cm}
\section{Conclusion}\label{sec:conclusion}
\vspace{-0.2cm}
We presented a blind calibration scheme for ULAs with coarsely quantized one-bit measurements, using a KLD-based calibration scheme. The KLD covariance fitting criterion---closely related to the optimally weighted nonlinear least squares criterion---implicitly exploit ``hidden" SOS information, thus leading to higher accuracy. We also provided an approximate iterative solution algorithm, along with an educated initial estimate (consistent in itself). Significant performance improvements, both in terms of calibration and a post-calibration estimation task, were demonstrated in simulations.


\bibliographystyle{IEEEbib}
\bibliography{refs}

\begin{thebibliography}{10}

\bibitem{walden1999analog}
Robert~H. Walden,
\newblock ``{Analog-to-Digital Converter Survey and Analysis},''
\newblock {\em IEEE J. Sel. Areas Commun.}, vol. 17, no. 4, pp. 539--550, 1999.

\bibitem{le2005analog}
Bin Le, Thomas~W. Rondeau, Jeffrey~H. Reed, and Charles~W. Bostian,
\newblock ``{Analog-to-Digital Converters},''
\newblock {\em IEEE Signal Process. Mag.}, vol. 22, no. 6, pp. 69--77, 2005.

\bibitem{sun2013wideband}
Hongjian Sun, Arumugam Nallanathan, Cheng-Xiang Wang, and Yunfei Chen,
\newblock ``{Wideband Spectrum Sensing for Cognitive Radio Networks: A
  Survey},''
\newblock {\em IEEE Wireless Communications}, vol. 20, no. 2, pp. 74--81, 2013.

\bibitem{stinco2014compressed}
Pietro Stinco, Maria Greco, Fulvio Gini, and Mario La~Manna,
\newblock ``{Compressed Spectrum Sensing in Cognitive Radar Systems},''
\newblock in {\em Proc. of ICASSP}, 2014, pp. 81--85.

\bibitem{smith2016experiments}
Graeme~E. Smith, Zach Cammenga, Adam Mitchell, Kristine~L. Bell, Joel Johnson,
  Muralidhar Rangaswamy, and Christopher Baker,
\newblock ``{Experiments with Cognitive Radar},''
\newblock {\em IEEE Aerosp. Electron. Syst. Mag.}, vol. 31, no. 12, pp. 34--46,
  2016.

\bibitem{choi2016near}
Junil Choi, Jianhua Mo, and Robert~W. Heath,
\newblock ``{Near Maximum-Likelihood Detector and Channel Estimator for Uplink
  Multiuser Massive MIMO Systems with One-Bit ADCs},''
\newblock {\em IEEE Trans. Commun.}, vol. 64, no. 5, pp. 2005--2018, 2016.

\bibitem{gao2017gridless}
Yulong Gao, Deshun Hu, Yanping Chen, and Yongkui Ma,
\newblock ``{Gridless 1-b DOA Estimation Exploiting SVM Approach},''
\newblock {\em IEEE Commun. Lett.}, vol. 21, no. 10, pp. 2210--2213, 2017.

\bibitem{ren2017one}
Jiaying Ren and Jian Li,
\newblock ``{One-Bit Digital Radar},''
\newblock in {\em Proc. 51st Asilomar Conference on Signals, Systems, and
  Computers}, 2017, pp. 1142--1146.

\bibitem{liu2017one}
Chun-Lin Liu and P.P. Vaidyanathan,
\newblock ``{One-Bit Sparse Array DOA Estimation},''
\newblock in {\em Proc. of ICASSP}, 2017, pp. 3126--3130.

\bibitem{ameri2019one}
Aria Ameri, Arindam Bose, Jian Li, and Mojtaba Soltanalian,
\newblock ``{One-Bit Radar Processing with Time-Varying Sampling Thresholds},''
\newblock {\em IEEE Trans. Signal Process.}, vol. 67, no. 20, pp. 5297--5308,
  2019.

\bibitem{huang2019one}
Xiaodong Huang and Bin Liao,
\newblock ``{One-Bit MUSIC},''
\newblock {\em IEEE Signal Process. Lett.}, vol. 26, no. 7, pp. 961--965, 2019.

\bibitem{ramamohan2019blind}
Krishnaprasad~Nambur Ramamohan, Sundeep~Prabhakar Chepuri, Daniel~Fern{\'a}ndez
  Comesana, and Geert Leus,
\newblock ``{Blind Calibration of Sparse Arrays for DOA Estimation with Analog
  and One-Bit Measurements},''
\newblock in {\em Proc. of ICASSP}, 2019, pp. 4185--4189.

\bibitem{loesch2013cramer}
Benedikt Loesch and Bin Yang,
\newblock ``{Cram{\'e}r-{Ra}o Bound for Circular and Noncircular Complex
  Independent Component Analysis},''
\newblock {\em IEEE Trans. Signal Process.}, vol. 61, no. 2, pp. 365--379,
  2013.

\bibitem{bar2002doa}
Ofer Bar-Shalom and Anthony~J. Weiss,
\newblock ``{DOA Estimation Using One-Bit Quantized Measurements},''
\newblock {\em IEEE Trans. Aerosp. Electron. Syst.}, vol. 38, no. 3, pp.
  868--884, 2002.

\bibitem{jacovitti1994estimation}
Giovanni Jacovitti and Alessandro Neri,
\newblock ``Estimation of the autocorrelation function of complex gaussian
  stationary processes by amplitude clipped signals,''
\newblock {\em IEEE Trans. Information theory}, vol. 40, no. 1, pp. 239--245,
  1994.

\bibitem{van1966spectrum}
J.~H. Van~Vleck and David Middleton,
\newblock ``{The Spectrum of Clipped Noise},''
\newblock {\em Proceedings of the IEEE}, vol. 54, no. 1, pp. 2--19, 1966.

\bibitem{petersen2008matrix}
Kaare~Brandt Petersen, Michael~Syskind Pedersen, et~al.,
\newblock ``The matrix cookbook,''
\newblock {\em Technical University of Denmark}, vol. 7, no. 15, pp. 510, 2008.

\bibitem{gray2006toeplitz}
Robert~M Gray,
\newblock ``Toeplitz and circulant matrices: A review,''
\newblock {\em Foundations and Trends{\textregistered} in Communications and
  Information Theory}, vol. 2, no. 3, pp. 155--239, 2006.

\bibitem{astely1999spatial}
{Ast{\'e}ly, David and Swindlehurst, A. Lee and Ottersten, Bjorn},
\newblock ``{Spatial Signature Estimation for Uniform Linear Arrays with
  Unknown Receiver Gains and Phases},''
\newblock {\em IEEE Trans. Signal Process.}, vol. 47, no. 8, pp. 2128--2138,
  1999.

\bibitem{weiss2020asymptotically}
Amir Weiss and Arie Yeredor,
\newblock ``{Asymptotically Optimal Blind Calibration of Uniform Linear Sensor
  Arrays for Narrowband Gaussian Signals},''
\newblock {\em IEEE Trans. Signal Process.}, vol. 68, pp. 5322--5333, 2020.

\bibitem{mackay2003information}
{MacKay, David J.C.},
\newblock {\em {Information Theory, Inference and Learning Algorithms}},
\newblock {Cambridge University Press}, 2003.

\bibitem{weiss2020blind}
Amir Weiss,
\newblock ``{Blind Direction-of-Arrival Estimation in Acoustic Vector-Sensor
  Arrays via Tensor Decomposition and Kullback-Leibler Divergence Covariance
  Fitting},''
\newblock {\em \textnormal{submitted to} IEEE Trans. Signal
  Process.\textnormal{; also:} ArXiv e-prints}, May 2020,
\newblock arXiv:2005.08318 [eess.SP], https://arxiv.org/abs/2005.08318.

\bibitem{jennrich1976newton}
Robert~I. Jennrich and P.F. Sampson,
\newblock ``{Newton-Raphson and Related Algorithms for Maximum Likelihood
  Variance Component Estimation},''
\newblock {\em Technometrics}, vol. 18, no. 1, pp. 11--17, 1976.

\bibitem{bonnans2006numerical}
Joseph-Fr{\'e}d{\'e}ric Bonnans, Jean~Charles Gilbert, Claude Lemar{\'e}chal,
  and Claudia~A. Sagastiz{\'a}bal,
\newblock {\em {Numerical Optimization: Theoretical and Practical Aspects}},
\newblock Springer Science \& Business Media, 2006.

\bibitem{collier2005fisher}
Sandra~L. Collier,
\newblock ``{Fisher {I}nformation for a {C}omplex {G}aussian {R}andom
  {V}ariable: {B}eamforming {A}pplications for {W}ave {P}ropagation in a
  {R}andom {M}edium},''
\newblock {\em IEEE Trans. Signal Process.}, vol. 53, no. 11, pp. 4236--4248,
  2005.

\bibitem{paulraj1985direction}
Arogyaswami Paulraj and Thomas Kailath,
\newblock ``{Direction of Arrival Estimation by Eigenstructure Methods with
  Unknown Sensor Gain and Phase},''
\newblock in {\em Proc. of ICASSP}, 1985, vol.~10, pp. 640--643.

\bibitem{weiss2019blindnumberofsources}
Amir Weiss and Arie Yeredor,
\newblock ``{Blind Determination of the Number of Sources Using Distance
  Correlation},''
\newblock {\em IEEE Signal Process. Lett.}, vol. 26, no. 6, pp. 828--832, 2019.

\bibitem{he2010detecting}
Zhaoshui He, Andrzej Cichocki, Shengli Xie, and Kyuwan Choi,
\newblock ``{Detecting the Number of Clusters in $n$-Way Probabilistic
  Clustering},''
\newblock {\em IEEE Trans. Pattern Anal. Mach. Intell.}, vol. 32, no. 11, pp.
  2006--2021, 2010.

\end{thebibliography}

\end{document}